\documentstyle[emulateapj,psfig]{article} \begin{document}

\title{Discovery of Pulsed X-rays from the SMC Transient RX J0052.1-7319}

\author{M. H. Finger\altaffilmark{1}}
\affil{National Space Science and Technology Center, ES 50,
320 Sparkman Drive, Huntsville , AL 35805.
\\mark.finger@msfc.nasa.gov} 

\author{D. J. Macomb\altaffilmark{1}}
\affil{Laboratory for High Energy Astrophysics, code 661, 
NASA/Goddard Space Flight Center, Greenbelt, MD 20771.
\\macomb@cossc.gsfc.nasa.gov}

\author{R. C. Lamb, T. A. Prince}
\affil{Space Radiation Laboratory, California Institute of Technology, 
Pasadena, CA 91125.
\\lamb@srl.caltech.edu,prince@caltech.edu}

\author{M. J. Coe and N.J. Haigh}
\affil{Dept. Of Physics \& Astronomy, The University, Southampton,
SO17 1BJ, England.\\mjc@astro.soton.ac.uk}

\altaffiltext{1}{Astrophysics Programs, Universities Space
Research Association}

\begin{abstract}
Coherent 65 mHz pulsations in the X-ray flux of the Small Magellantic Cloud
(SMC) transient source RX J0052.1-7319 have been detected by us in an 
analysis of ROSAT data. We report on the pulsations we detected in 
ROSAT HRI data and simultaneous detection of these pulses in hard X-rays 
using BATSE data. The BATSE data show an outburst of the source lasting 60
days. We report on optical observations of the candidate companion, and a
new source position we determined from the HRI data, which is consistent 
with the candidate's location. From the measured fluxes and observed 
frequency derivatives we exclude the possiblity that the pulsar is in the
foreground of the SMC, and show that an accretion disk is present during 
the outburst, which peaked near Eddington luminosity. 

\end{abstract}


\section{Introduction}

RX J0052.1-7319 is an X-ray source located in the the Small Magellanic Cloud,
first detected with Einstein (1E 0050.3-7335, \cite{Wang92}). 
It was classified by Kahabka and Pietsch (1996\markcite{Kahabka96}) as a 
transient X-ray binary candidate based on the ROSAT PSPC data from 
October 1991 and April 1992.

As part of a systematic search of the ROSAT data for pulsed sources, we
have discovered pulsations from RX J0052.1-7319 at a 
frequency of 65.4 mHz (period = 15.3 s)
using ROSAT HRI observations from 1996 November-December.
Our search had previously detected another SMC pulsar,
J0117.6-7330 (\cite{Macomb99}). 

After a preliminary report of our discovery of the pulsar nature of RX
J0052.1-7319 (\cite{Lamb99}), Kahabka estimated a new source position based on 
HRI observations of 1995 May (\cite{Kahabka99a}), and confirmed the detection
of pulsations using an HRI observations from October 1996 (\cite{Kahabka96}).
Israel et al. (1999\markcite{Israel99}) identified a Be star as the likely
optical counterpart to the source. Udalski (1999\markcite{Udalski99}) 
then reported on the long term optical variability of this candidate, 
based on Optical Gravtational Lensing Experiment (OGLE) monitoring.   

Here we report on the pulsations we detected in the ROSAT HRI data, detection
of these pulses in hard X-rays using simultaneous BATSE data, the history
of the outburst visible in the BATSE data, and optical observations of the
candidate companion and nearby stars.

\section{Observations}

We initially discovered pulsations from RX J0052.1-7319 in observations made
with the High Resolution Imager (HRI) detector of the ROSAT (\cite{Trumper83}) 
in 1996 November and December. The HRI, which consisted of two cascaded
microchannel plates (MCPs) with a crossed grid position readout system, was
sensitive to X-rays in the $\sim$ 0.2 -- 2 keV range.  Thereafter we detected
the pulsations in observations from the Burst and Transient Source Experiment
(BATSE) (\cite{Fishman89}) on the CGRO using data from the Large Area Detectors
(LADs) which are NaI scintillation detectors 
sensitive to hard X-rays/soft gamma-rays in the 20 keV to 2
MeV range.
 
\subsection{ROSAT Observations}
Our discovery was made in the ROSAT HRI observations of 1996 November 10.71 --
December 9.16 as part of a systematic search of the ROSAT data for previously
undetected pulsars.  These data were among 1365 data sets we selected from  the
catalog of 59911 ROSAT/HRI point source observations based on the  potential
for detection of significant pulsations. Each data set was processed using 
standard Ftools, with barycentered arrival times for the first 200 ks of each
observation binned with 5 ms resolution, and Fourier transformed. The power
spectra were searched for significant pulsed signals unassociated with the
$\sim$5760 s spacecraft orbit or the 402 s period spacecraft orientation
wobble. The RX J0052.1-7319 HRI observations were selected for further analysis
because of a strong signal near 65 mHz.

\begin{minipage}{3.3in}
\psfig{file=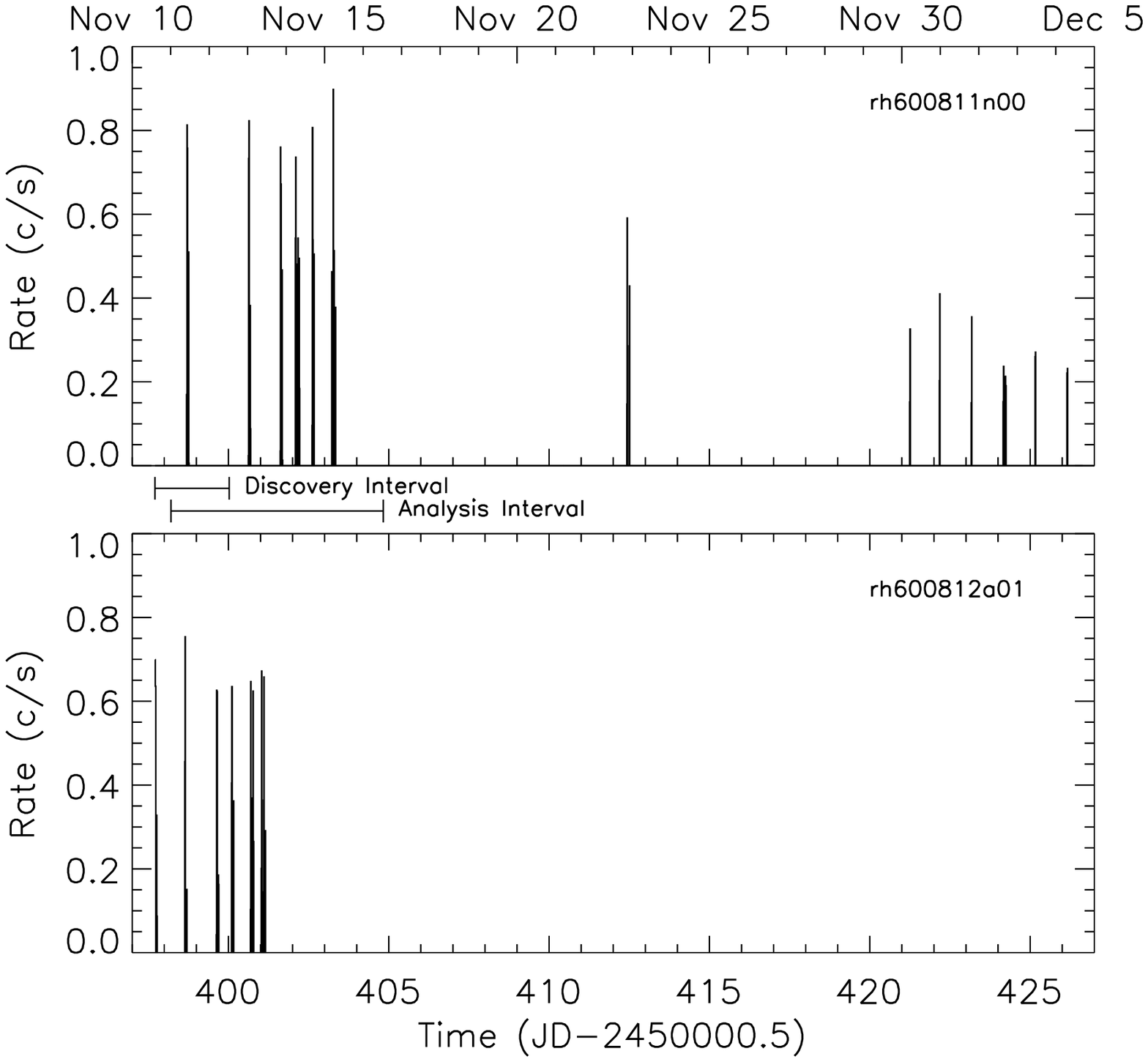,width=3.25in}
\figcaption[f1.ps]{ 
Mean count rates in 1000 s bins for the 1996 November 11 (MJD 50398) --
December 9 ROSAT HRI observations of RX J0052.1-7319 showing the time structure
of the observation window, and the declining flux of the source. The upper
panel is for observation rh600811n00 and the lower for rh600812a01. Between the
panels is shown both the time interval for the initial FFT in which 
the pulsations
were discovered, and the interval used for the analysis resulting figures 2 and
3.\label{HRI_rates}}
\vspace*{10pt}
\end{minipage}

These HRI observations, rh600811n00 and rh600812a01, are of two SMC fields
which contain RX J0052.1-7319. The observations for these two fields are
partially contemporaneous.  Figure 1 shows the count rates in 1000 s bins for
events  within $60\arcsec$ of the source in these observations.
The roughly 10-20\%
difference in counting rates between contemporaneous portions of the
two observations may be due principally to differences in source
vignetting.  For observation rh600812a01, the source is $17\arcmin$
from the center of the HRI field and vignetting corrections are
10-20\%; for observation rh600811n00 the source is $8\farcm 4$ from
the HRI center and vignetting is substantially less.  
From observation rh600811n00 it is clear that the source flux is declining.

As can be seen in Figure 1,
the initial 200 ks (2.3 days) interval of the observations, 
which was used for the initial search fft, contained
only a portion of either of the two observations. We chose for
further analysis the data of Nov 11.21-16.83 (all of rh600812a01 and
the first portion of rh600811n00).  

From initial analyses of this dataset we knew pulse frequency was changing 
significantly, and that there was substantial power at higher harmonics of 
the pulse frequency. To accurately estimate the pulse frequency and 
frequency derivitive we maximized the $Z^2_3$ statistic 
(\cite{Buccheri83}),
where $Z^2_3(f) = p(f) + p(2f) + p(3f)$, and the Rayleigh statistic
$p(f)$ is calculated as:
\begin{equation} 
  p(f) = {2 \over N} \left|\sum_{k=1}^N 
         exp(i2\pi f [t_k+\onehalf\alpha (t_k-\tau)^2])\right|^2~. 
\end{equation}

\begin{minipage}{3.3in}
\psfig{file=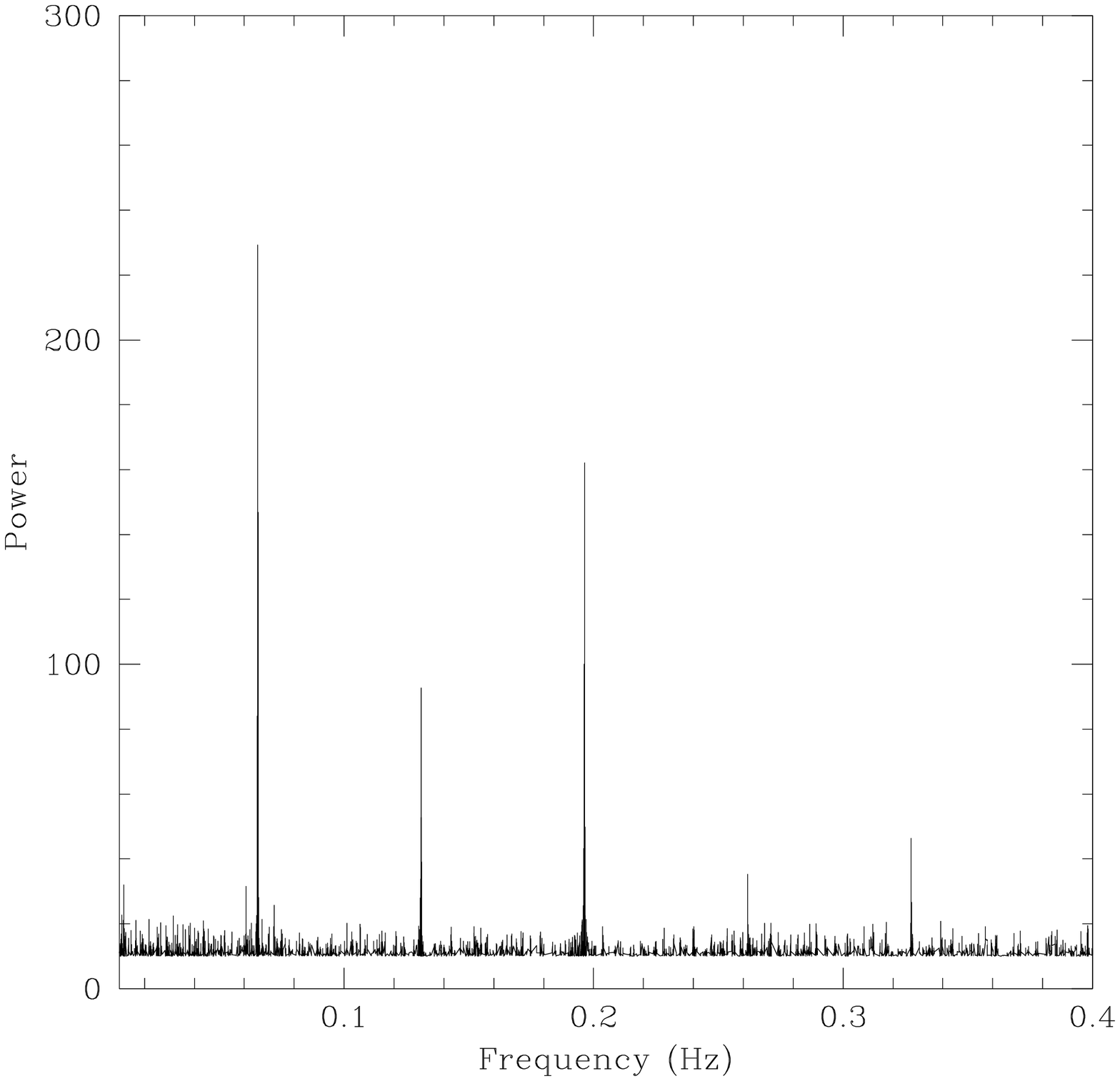,width=3.25in}
\figcaption[f2.ps]{ 
Rayleigh periodogram for the 1996 Nov 11.21-16.83 ROSAT HRI observations
of RXJ 0052.1-7319. This was implemented with an FFT of data binned with
0.5\,s resolution. A fractional frequency derivative of 
$\dot\nu/\nu = 8.37\times 10^{-10}$~s$^{-1}$ is accounted for in the 
analysis. Only values above 10 are plotted. 
The pulse frequency, near 65 mHz, and four harmonic overtones 
are clearly detected. The structure in the wings of each peak is due to the
non-uniform spacing of the data.\label{RX_periodogram}}
\vspace*{10pt}
\end{minipage}

Here $N$ is the number of photons detected in the interval, 
$f$ is the analysis frequency, $\alpha = \dot\nu / \nu$, with $\nu$ the
pulse frequency, $t_k$ is a barycentric arrival time of photon $k$, 
and $\tau$ is an epoch within the data interval.

For epoch MJD 50401.0 (1996 Nov. 14.0) we obtained $\nu$ = 0.06545850(7) Hz and
$\dot\nu = 5.48(7)\times 10^{-11}~{\rm Hz}~{\rm s}^{-1}$. The maximum value
of  $Z^2_3$ was 524, which is highly improbable by 
chance.\footnote[1]{We estimate that the probability of
obtaining a $Z^2_3$ of 524 or more due to Poisson noise is 
less than $10^{-98}$, including the number of trials introduced by searching
in frequency and frequency derivative. This calculation however neglects
systematic signatures in the data, which would likely play a dominate role
any false detection with a $Z^2_3$ this large.}  
The Rayleigh statistic for this ratio of $\dot\nu/\nu$ 
is shown in figure \ref{RX_periodogram}. 
In addition to the pulse fundamental, four pulse harmonic
overtones are clearly detected.

The HRI data from the 1996 November
11.21-16.83 interval epoch-folded with this ephemeris is shown in
figure \ref{HRI_profile}.  The pulse profile has a single 
asymmetric peak, with a narrow valley at minimum.  The pulsed fraction
([mean-minimum]/mean) is $35\pm3$\%.

For the HRI data from the interval 1996 December 5-9 we obtain a frequency 
$\nu = 0.0655330(3)$\,Hz at epoch MJD 50524.0, and  
$\dot\nu = 2.3(2)\times 10^{-11}~{\rm Hz}~{\rm s}^{-1}$.  
Thus the decrease in counting rate noted
above was accompanied by a decrease in spin-up rate, as would be expected
from the intrinic correlation between mass accretion rate and angular momentum
accretion rate in a disk fed accreting pulsar.

For the 1996 November 11-16  and December 5-9 intervals we find source count
rates of 0.92 and 0.46 counts\,s$^{-1}$ respectively. These rates are corrected
for vignetting (\cite{HRI_CAL}) and the  $15\arcsec$ radius circle used to
select the data, using the HRI  off-axis encircled energy profile
(\cite{Boese00}). 

\begin{minipage}{3.3in}
\psfig{file=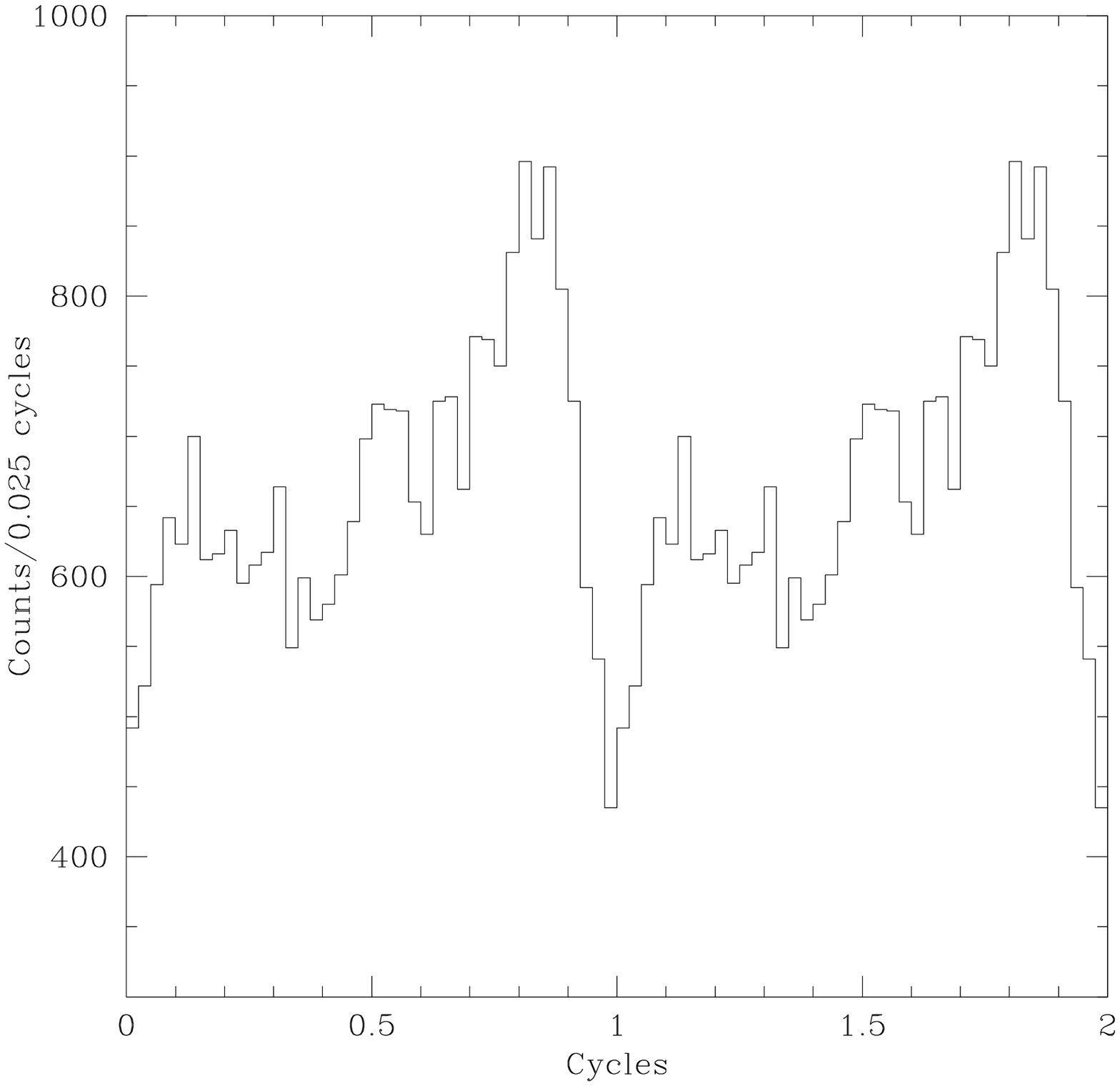,width=3.25in}
\figcaption[f3.ps]{
Pulse profile of RXJ 0052.1-7319 from the ROSAT HRI observations of 1996 
Nov 11.21-16.83.\label{HRI_profile}}
\vspace*{10pt}
\end{minipage}

To estimate fluxes from these count rates we have used the
spectrum estimated for the source by Kahabka \& Prietsch (1996) based on a fit
of ROSAT PSPC data. This consists of thermal bremsstrahlung model with  $kT=17$
keV, a galactic absorption column of $3\times10^{20}{\rm cm}^{-2}$,  and an
absorption column within the SMC with density of $N_H = 7.0\pm4.0 \times
10^{21}  {\rm cm}^{-2}$ (95\% confidence errors), with metallicities reduced by
a factor of seven from solar system abundances. This model results in 0.1-2.0
keV unabsorbed flux estimates of $5.9\pm1.0 \times 10^{-11} {\rm erg}~{\rm
cm}^{-2} {\rm s}^{-1}$ and of $3.0\pm0.5 \times 10^{-11} {\rm erg}~{\rm
cm}^{-2} {\rm s}^{-1}$ for the 1996 November 11-16 and December 5-9 intervals
respectively.

Using the 1996 November data of observation rh600811n00, we
have determined the position of the source to be 
R. A. = $0^{\rm h}52^{\rm m}13\fs 65$,
Dec. = $-73\arcdeg 19\arcmin 19\farcs 5$, 
with a statistical uncertainty ($1\arcsec$) which is
negligible in comparison to systematic position errors,
which result in an 68\% confidence error circle of radius
7$\arcsec$ (\cite{Kurster93},
ROSAT Users Handbook\footnote[2]{ 
http://heasearc.gsfc.nasa.gov/docs/rosat/ruh/handbook/node34.html}). 
These systematic errors are due to uncertainty in the aspect of ROSAT.
We note that this position differs by nearly 10" from the position quoted
by Kahabka (1999b\markcite{Kahabka1999a}) and Kahabka
(2000\markcite{Kahabka00}).  However, since
these latter determinations used either very low rate data ($<$0.01
c/s) or data in which the source was at the edge of the HRI field of
view, we believe that the position given above may be the more accurate.

\subsection{BATSE Observations}

The Burst and Transient Source Experiment (BATSE) was an all sky monitor  which
flew onboard the Compton Gamma-Ray Observatory (CGRO). The data used in the
analysis presented here were from the Large Area Detectors (LADs) which were 
NaI(Tl) scintillation counters 1.27 cm thick, with 2025 cm$^2$ area, that were
located at the eight corner of the CGRO spacecraft.  Count rate data from the
LADs was continuously available from 1991 April to 2000 May, except during SAA
passages and occasional telemetry outages.

The 20-50 keV channel of the BATSE LAD discriminator rates (DISCLA channel 1)
were analysed for pulsations from RX J0052.1-7319 using techniques discussed in
Finger et al. (1999\markcite{Finger99}). 
Rates were combined from different
detectors with coefficients optimal for a source with a spectrum of the form
$dN/dE = A*exp(-E/kT)/E$ with $kT = 20$ keV. A large number of pulse profiles
are obtained by fitting short segments of these combined rates with a model
consisting of a quadratic spline background, plus a low order Fourier expansion
pulse profile model. These profiles are then combined over multiple day
intervals using trial frequencies and frequency rates. The resulting combined
profiles are evaluated with the $Y_n$ statistic (\cite{Finger99}), 
which is simular to the $Z^2_n$ statistic, 
but accounts for possible non-Poisson noise.

Using data from the interval 1996 Nov 11.0--17.0, $Y_3$ was searched over a
frequency range of width $10^{-4}$ Hz centered on the ROSAT measurement, 
and a frequency rate range of $ -10^{-11}~{\rm Hz}~{\rm s}^{-1}$ to 
$10^{-10}~{\rm Hz}~{\rm s}^{-1}$ resulting in maximum value of 
$Y_3 = 57.0$ which we
estimate\footnote[3]{This estimate approximates the
distribution of $Y_3$ with the chi-squared distribution with 6 
degrees of freedom, which is
valid in this case because of the large number of profiles being combined,
and accounts for 1100 independent trials.} 
has a probability of being exceeded by chance of  less than 
$2\times 10^{-7}$. The resulting frequency and frequency rate estimates 
were $\nu = 0.06545830(10)$~Hz at epoch MJD 50401.0, and 
$\dot\nu = 5.62(12)\times 10^{-11}$~Hz~s$^{-1}$, in
good agreement with the ROSAT results. 

The BATSE pulse profile for this interval using the ROSAT HRI ephemeris is
shown in figure \ref{BATSE_profile}. The profile is shown relative to the mean
flux level, which cannot be determined from the data due to the weakness of the
source and the high background level. 
The solid curve is the profile corresponding 
to six Fourier coefficients which have been estimated with 
the same fitting technique
used with the frequency search.
Error bars for the
value of the curve at 13 approximately independent points are shown. Six
harmonics were chosen because this approximately matches the 1.024 s resolution
of the data.

Pulsations were then detected in nine additions six-day intervals, extending
the total period the source was detected with BATSE to 60 days between 1996
September 18 and November 17. Beginning with the Nov 11-17 interval, the total
period of detection was progressively extended outward by frequency and
frequency rate searches in the adjacent six day intervals, using $5\times
10^{-5} Hz$ frequency ranges centered on the extrapolation from the neighboring
frequency and frequency rate estimate, and the frequency rate range given
above. Searching in this manner at the boundaries of the known frequency
history reduces the required frequency search range, providing the best
sensitivity. Figure \ref{BATSE_timing} shows the resulting measurements of
frequency, frequency rate,  and r.\,m.\,s. pulsed flux in the 20-50 keV energy 
band. The figure also shows for comparison the frequency and frequency rate 
measurements from the ROSAT HRI data.

\begin{minipage}{3.3in}
\psfig{file=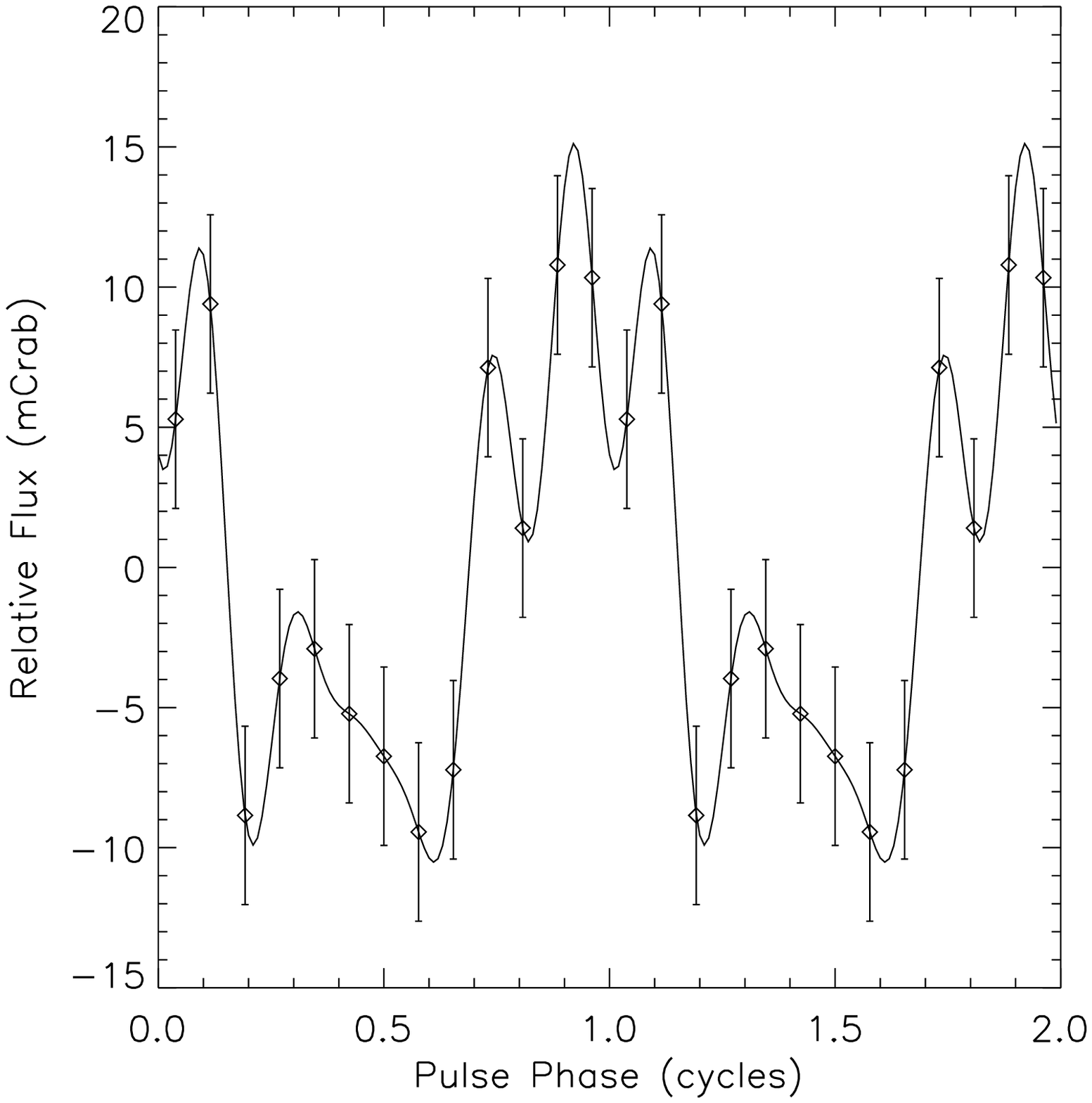,width=3.25in}
\figcaption[f4.ps]{
Pulse profile of RXJ 0052.1-7319 from the BATSE observations of 1996 
Nov 11.0-17.0. This profile is based on an estimated six harmonic Fourier
model. Error bars are given for approximately independent points.
This profile is folded with the same ephemeris at that in figure 3.
\label{BATSE_profile}}
\vspace*{10pt}
\end{minipage}

Efforts were made to detect the source from 1996 August 1 to September 18, and
from 1996 November 17 to 1997 January 8, but these failed.  These attempts used
searches with a frequency range of $3\times 10^{-4} $Hz centered on the nearest
detection, and a frequency rate range of $ -10^{-11}~{\rm Hz}~{\rm s}^{-1}$ to
$10^{-10}~{\rm Hz}~{\rm s}^{-1}$. Prior to 1996 September 18, 
pulsations from the
130.4 mHz pulsar 4U 1626-67 interfered with  first harmonic (i.e. $2\times f$) 
contribution to $Y_3$, in a narrow frequency band  near 65.2\,mHz. For this
narrow band, this harmonic was left out of  the search. We conclude that in
these intervals prior to and following our 1996 September 18 to  November 17
detections that the source had pulsed flux below our  sensitivity level. We
estimate an upper limit of 8 mCrab for the 20-50 keV r.\,m.\,s. pulsed flux for
these intervals. 

\subsection{Optical Observations}
 
In figure \ref{v2} we show a V band image taken from South African Astronomical
Observatory (SAAO) 1.0m telescope on 20 Jan 1999.
Marked on the figure is the X-ray error circle of 
Kahabka (2000\markcite{Kahabka00}) which has a 6" radius, and the
error circle from this work, with a 7" radius. 
The object marked A is the source identified as the counterpart by 
Israel et al. (1999\markcite{Israel99}) based upon a red spectrum. 
The objects labeled A, B, C, and D are also identified in Figure \ref{ha_r}.

If the optical counterpart of RX J0052.1-7319 is a Be star, then it should show
excess $H_\alpha$ emission. In figure \ref{ha_r} we show $H_\alpha$  versus R
band counts for 25 stars in  the vicinity of Kahabka's X-ray error circle.
The line is a linear best fit through all the data points shown. 
Stars that lie above the line are  ones that show an $H_\alpha$ excess -- i.e.
objects A, B and C as identified on figure \ref{v2}.  
Object D, the one inside the Kahabka (2000\markcite{Kahabka00})
error circle, shows no excess. Objects B and C, though exhibiting an
$H_\alpha$ excess, are too far away from the X-ray position to be the
counterpart. If, as we might expect, the optical companion is a Be star,
the only reasonable counterpart is object A, the star identified by Isreal et
al. (1999). Object A is well contained with the error circle for RX
J0052.1-7319 presented here, and thus the identification of the X-ray source
as a member of a Be system seems reasonable.

The absolute V magnitude of object A may be determined from the value quoted by 
Udalski (1999\markcite{Udalski99}) of $m_v$=14.67. The distance modulus to the 
SMC determined by Westerlund (1997\markcite{Westerlund})
is $(m-M)_o$ = 18.9. In addition, the average extinction is E(B-V) =
0.07-0.09 (\cite{Schwering91}), though there are 
regions in the SMC where it can rise as high as 0.25. Combining these
parameters leads to an estimate of $M_v$= -4.47$\pm$0.02. This magnitude
corresponds to a star in the range B1III - B0V, very similar to the 
value obtained for Be/X-ray binary counterparts (e.g. that of 
RX J0117.6-7330 quoted in Coe et al. (1998\markcite{Coe98})).
\vspace{1.0in}

\begin{minipage}{3.3in}
\psfig{file=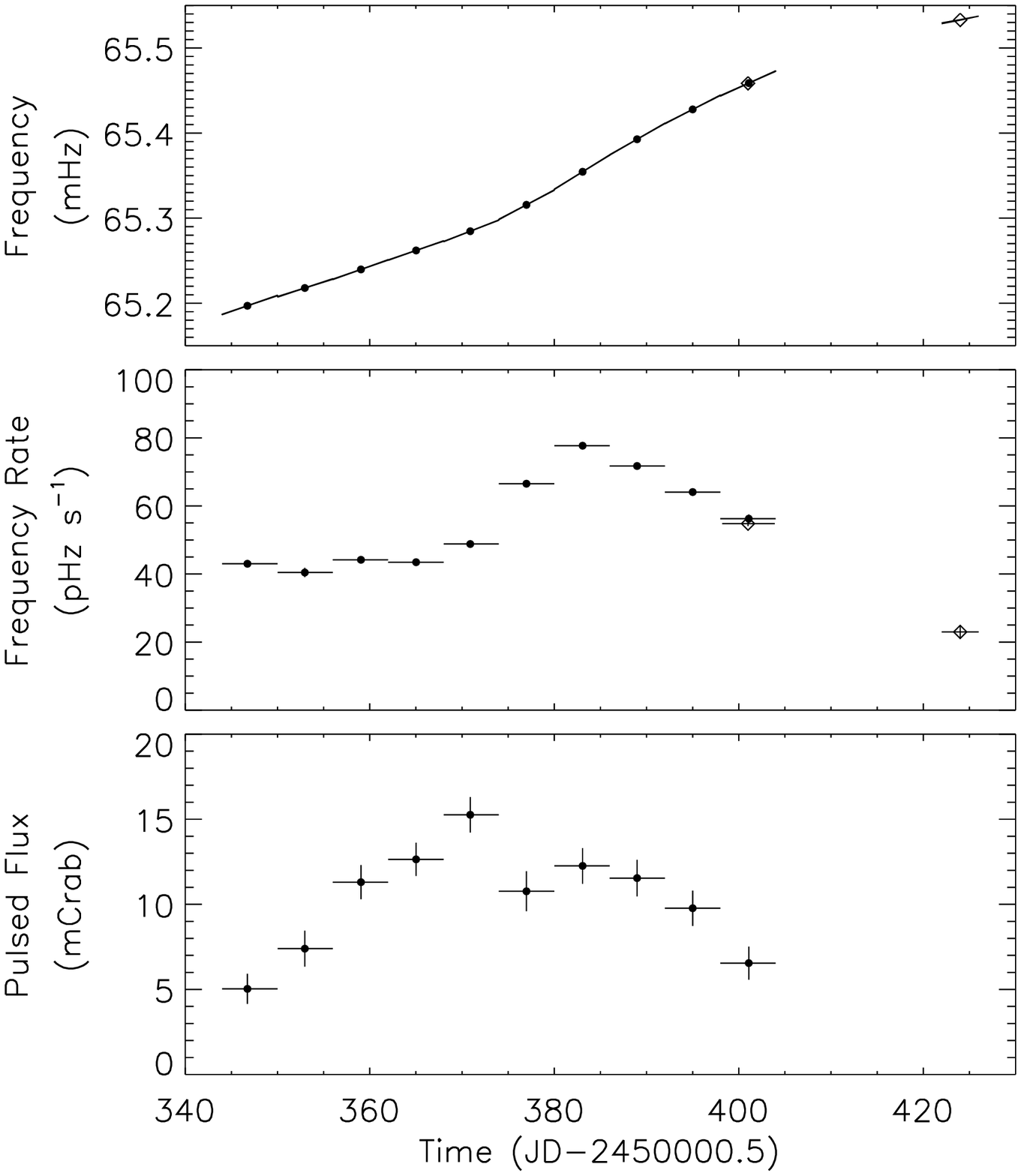,width=3.25in}
\figcaption[f5.ps]{Pulse timing measurement for RX J0052.1-7319
from BATSE and ROSAT, and pulsed flux measurements from BATSE. 
The top panel shows the pulse frequency, with 
solid circles for BATSE measurements, and open diamonds for 
ROSAT HRI measurements.
The lines through each point span the measurement interval, and have the slope
of the measured frequency rate. The error bars are too small to be visible
on the plot.
The middle panel shows the pulse frequency rate, with 
solid circles for BATSE measurements, and open diamonds for 
ROSAT HRI measurements. The lines through each point span the measurement
interval. The bottom panel shows the BATSE r.\,m.\,s. pulsed flux (20-50 keV).
For reference 1 mCrab (20-50 keV) = 
$1.0\times 10^{-11}~{\rm erg}~{\rm cm}^2~{\rm s}^{-1}$, which corresponds to a
luminosity of $4.3\times 10^{36} {\rm erg}~{\rm s}^{-1}$ in the SMC.
The plot covers the date range 1996 September 14 to December 23. For flux upper
limits before and after the BATSE measurements presented see the
text.
\label{BATSE_timing}}
\vspace*{10pt}
\end{minipage}

\begin{minipage}{3.3in}
\psfig{file=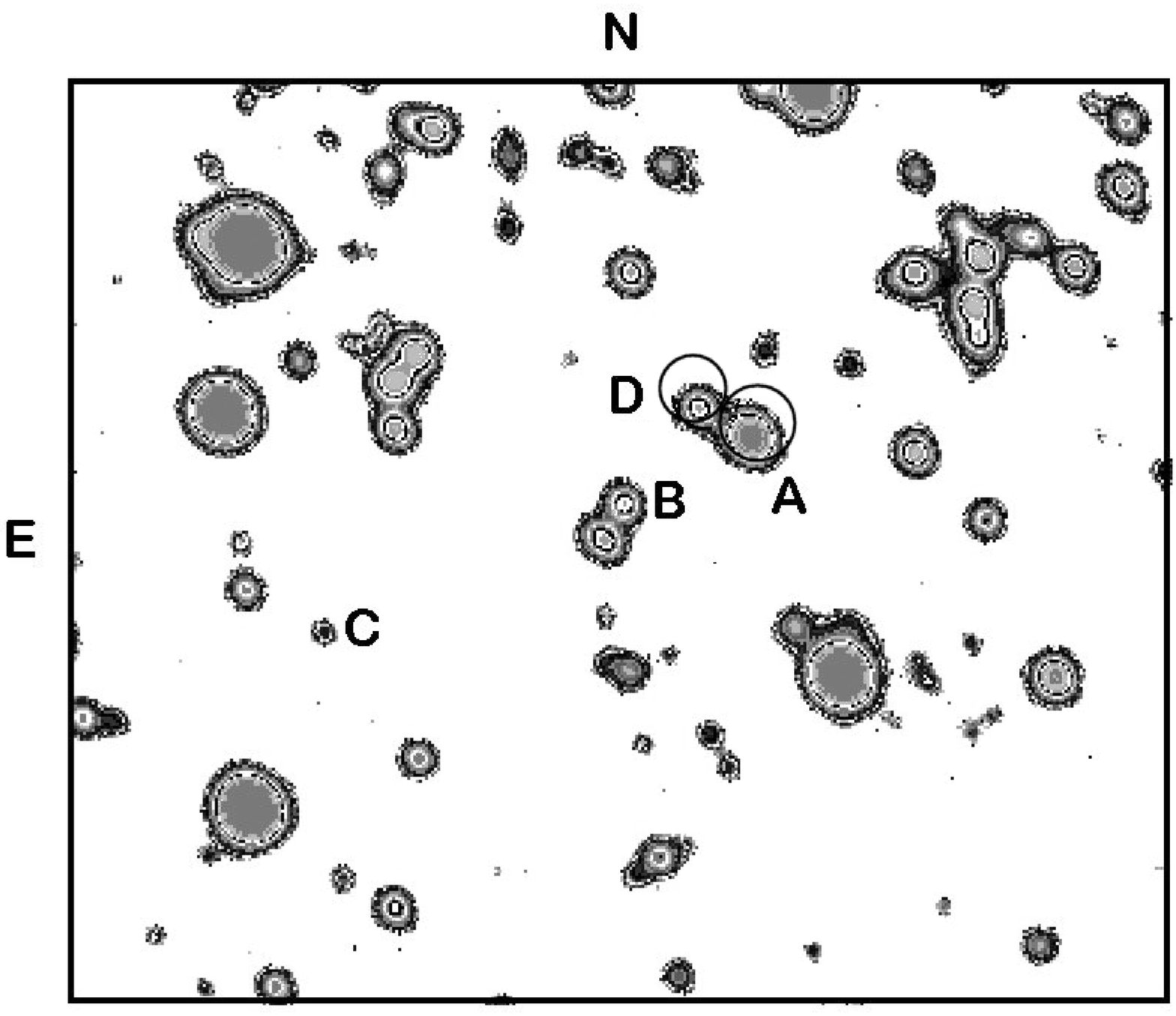,width=3.25in}
\figcaption[f6.eps]{
V band image taken from South African Astronomical Observatory (SAAO) 1.0m
telescope on 20 January 1999. Marked on this 
figure is the X-ray error circle reported in Kahabka (2000) near Star D, 
and the error circle from this work located 
near Star A.\label{v2}}
\vspace*{10pt}
\end{minipage}

\section{Discussion}

The ROSAT HRI observations we have presented show that RX J0052.1-7319 is a
pulsar with a 65 mHz rotation frequency. Variations in the flux by more than 
two orders of magnitude between ROSAT observations at different epochs  
demonstrates that the source is a transient (\cite{Kahabka96},\cite{Lamb99}).
The BATSE observations we have presented show that the November -- December
1996 ROSAT HRI observations occurred on the tail of a outburst that began at
least two months earlier, lasting more than 80 days. 

The companions of transient pulsars with known spectral type  are generally Be
(or Oe) stars. There are now known a few transient pulsars in low-mass X-ray
binaries, but these all have high spin frequencies  ($\nu > 1$ Hz). Be stars
show both emission in the Balmer lines, and an excess in the IR, due to a disk
of material shed from the equator of the rapidly rotating star. The accretion
of this material, which is thought to form a quasi-keplerian disk, fuels
the X-ray outbursts.

Two types of outbursting behavior are observed in Be/pulsar binaries: type I
(``normal" outbursts), with a series of lower luminosity  ($L < 10^{37}~{\rm
erg}~{\rm s}^{-1}$ outbursts occurring once per orbit, generally near
periastron; and type II (``giant" outbursts), single high luminosity 
($L\sim 10^{38}~{\rm erg}~{\rm s}^{-1}$ sometimes lasting several orbits 
(\cite{Stella86}). These outbursts are generally accompanied by rapid spin-up
of the neutron star, indicating disk accretion (\cite{Finger96,Bildsten97}).

The BATSE observations presented here are consistent with a giant outburst
of a Be/pulsar transient in the SMC. From the observed spin-up we can show
that a disk is present, and the the source is outside our galaxy. In disk
accretion, the angular momentum accreted per unit accreted mass is 
$l = (GMr_m)^{\onehalf}$, 
were $M$ is the neutron star mass, and $r_m$ the radius
of the magnetosphere.\linebreak

\begin{minipage}{3.3in}
\psfig{file=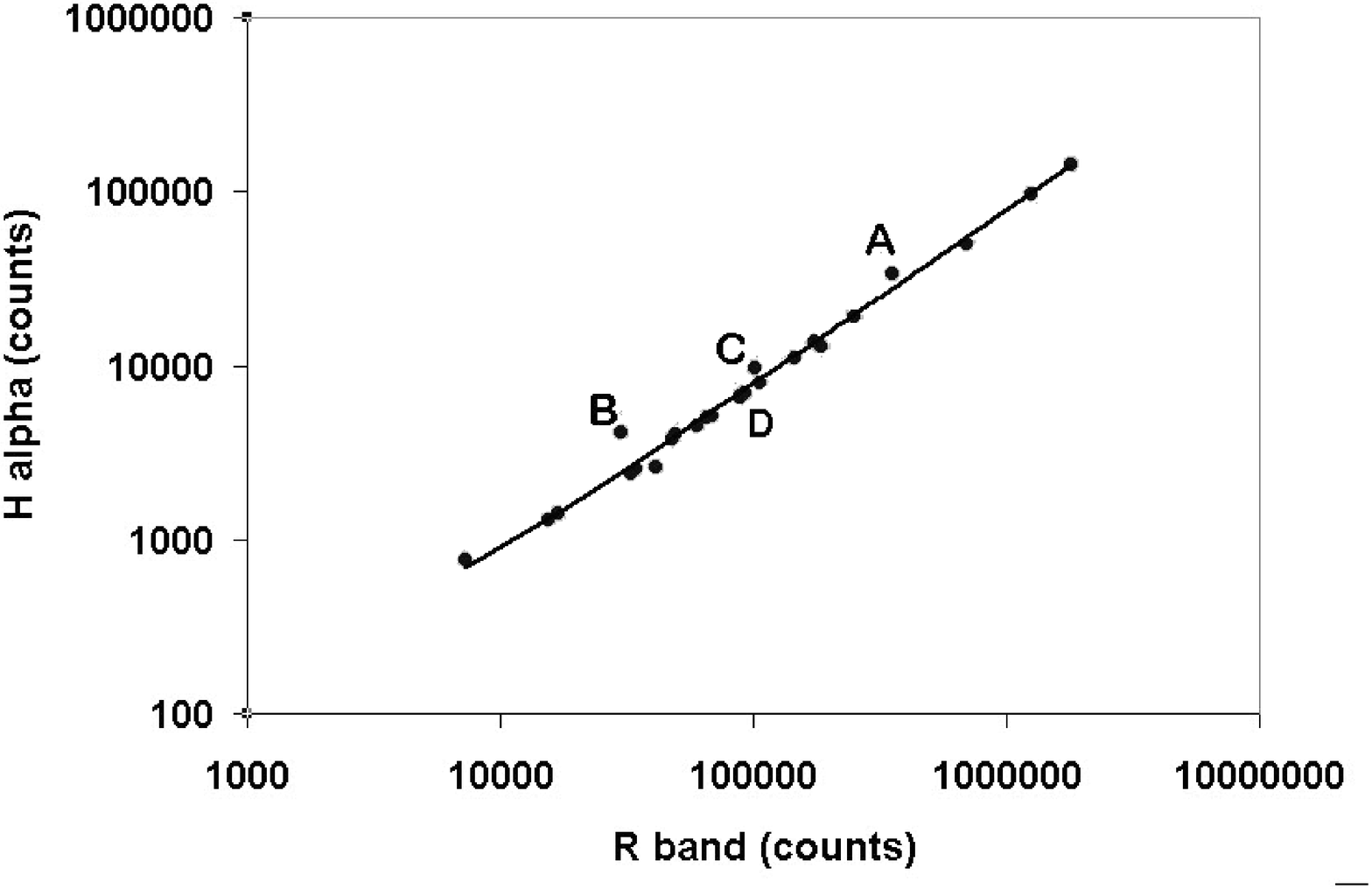,width=3.25in}
\figcaption[figure_7.eps]{$H_\alpha$ versus R band plot of 25 stars in the 
vicinity of the X-ray error circle. Stars that lie above the line are ones 
that show an H alpha excess. Letters refer to stars identified in
the previous figure.\label{ha_r}}
\vspace*{10pt}
\end{minipage}

\noindent The specific angular momentum for wind accretion is
generally much less than this, since if it reaches this value, a disk forms.
The magnetospheric radius is limited by the corotation radius 
$r_c = (GM)^{\onethird} (2\pi\nu)^{-\twothirds} = 1.0\times 10^9~{\rm cm}$, 
where we have used $M = 1.4 M_\odot$. The specific angular momentum $l$ is 
proportional to the ratio $\dot\nu / L$ where $L$ is the pulsar's luminosity. 
Since the rate at which angular momentum is accreted is $l \dot m = 2 \pi I
\dot \nu$ where I is neutron star moment of inertia, and $L = G M \dot
m/R$, where $R$ is the neutron star radius, we have the upper limit
\begin{equation}
{{\dot\nu} \over L} < (2\pi)^{-\case{4}{3}}(GM)^{-\case{1}{3}}
  I^{-1} R \nu^{-\case{1}{3}} = 3.8\times 10^{-49}~{\rm Hz}~{\rm erg}^{-1}
\label{eqn2}
\end{equation}
where we have used a moment of inertia $I=10^{45} {\rm gm}~{\rm cm}^2$,
and a radius $R=10^6~{\rm cm}$. 
For the Nov 11-17 observation, for which $\dot\nu=5.6 \times 10^{11} {\rm\ Hz\
s}^{-1}$, this implies
$L > 1.5\times 10^{38}~{\rm erg}~{\rm s}^{-1}$, and for peak frequency rate
observed we have $L > 2.0\times 10^{38}~{\rm erg}~{\rm s}^{-1}$. Using the
0.1-2.0 keV flux $F=5.9\times 10^{-11}~{\rm erg}~{\rm cm}^{-2}~{\rm s}^{-1}$ 
observed in the 1996 Nov 11-17 HRI observation, we find
the lower limit on the source distance of
\begin{equation}
 d > 46\left({{\epsilon} \over {0.1}}\right)^{\case{1}{2}}~{\rm kpc}
\end{equation}  
where $\epsilon$ is the fraction of the luminosity in the 0.1-2.0 keV energy
band. This eliminates the possibility that the RX J0052.1-7319 is in the
foreground of the SMC. 

The limit in equation \ref{eqn2} is reached only with disk accretion. Using the
measured frequency rate, fluxes, and the distance of 60 kpc to the SMC we find
a ratio of ${{\dot\nu}/L}$ comparable to this limit, suggesting that an
accretion disk is present during the outburst.

The pulse periods of Be binary pulsars range from 69\,ms to 1400\,s, with
determined orbital periods ranging from 17 to 250 days.  Corbet
(1986\markcite{Corbet86}) showed that the orbital period is correlated with the
pulse period. From the observed distribution of spin and orbital periods we
would expect an orbital period in the range of 25 -- 100 days.  In giant
outbursts of Be/X-ray pulsars we expect a strong correlation of flux and
frequency rate (see e.g. \cite{Finger96}). We note however that the history of
the frequency rate in fig \ref{BATSE_timing} is  dissimilar in profile to that
of the pulsed flux. This could be due to the doppler signature of a
binary orbit. An alternate
explanation would be changes in the spectra or pulse fraction. However, 
giant outbursts typically have simple flux and intrinsic spin-up rate 
profiles, with a steady rise to peak, and a somewhat slower fall 
(see e.g. \cite{Parmar89,Whitlock89,Finger96}, but also, 
\cite{Negueruela97}).  

The plausible identification of RX J0052.1-7310 as Be binary system in
the SMC accentuates further the rather dramatic difference between the
SMC and our Galaxy with regard to the population of high mass X-ray
binaries.  This fact has already been noted by several authors
(\cite{Schmidtke99,Jokogawa00}).
A recent compilation of the known X-ray 
pulsars\footnote[4]
{http://gammaray.msfc.nasa.gov/batse/pulsar/asm\_pulsars.html} 
gives within the SMC one supergiant system, three known Be systems, and 11
transients with uncertain companion class (likely to be Be
systems), making 15 high mass X-ray pulsar binaries.  For the Galaxy the
corresponding number is 40.  Therefore, using a mass ratio of the SMC
to the Galaxy of 1/100, this suggests that high mass X-ray pulsar systems 
in the SMC are overabundant by roughly a factor of 30 relative to the Galaxy. 
This analysis ignores the important effects of obscuration within the Galaxy,
the relative frequency of observations, and the low luminosity sensitivites
obtained for the Magellantic clouds; nevertheless the apparent disparity is
remarkable.
 
Since high mass X-ray binaries have lifetimes which are a very small fraction
($\sim 10^{-3}$) of the age of the Galaxy, the dramatic difference between the
SMC and the Galaxy points to a rather recent outburst of star-formation in the
SMC within the last $\sim 10^7$ years. Further support of such an epoch of star
formation comes from the radio observations of H1 by Stavely-Smith et al. 
(1997\markcite{Stavely-Smith97}) and Putman et al. (1998\markcite{Putman98}),
which show a strong bridge of material between the Magellanic Clouds and
between them and our own galaxy. Furthermore, Stavely-Smith et al. have
demonstrated the existence of a large number of supershells
(created by multiple supernovae) of a
similar age ($\sim$5 Myr), strongly suggesting enhanced starbirth has taken
place as a result of tidal interactions between these component systems.
Consequently it seems very likely that the previous closest approach of the SMC
to the LMC $\sim 10^8 $ years ago may have triggered the birth of many new
massive stars which have given rise to the current population of HMXBs. In
fact, other authors (e.g. \cite{Popov98}) claim that the presence of large
numbers of HMXBs may be the best indication of starburst activity in a system.

M. H. F. acknowledges support through NASA grant NAG5-4238.

\end{document}